\documentclass[doublecol]{epl2}

\usepackage{graphicx}
\usepackage[usenames]{color}
\newcommand{\ket}[1]{\left|#1\right\rangle}
\newcommand{\bra}[1]{\left\langle#1\right|}
\newcommand{\vecbf}[1]{\mbox{\boldmath{$#1$}}}

\title{Spin-density induced by
electromagnetic wave in two-dimensional electron gas}
\shorttitle{Spin-density induced by electromagnetic wave in 2DEG}

\author{Alexander Shnirman\inst{1} and Ivar Martin\inst{2}}
\shortauthor{A. Shnirman and I. Martin}

\institute{
  \inst{1} Institut f\"ur Theoretische Festk\"orperphysik,
           Universit\"at Karlsruhe, D-76128 Karlsruhe, Germany.\\
  \inst{2} Theoretical Division, Los Alamos National Laboratory,
           Los Alamos, NM 87545, USA.
}

\pacs{72.25.Rb}{Spin relaxation and scattering}
\pacs{85.75.-d}{Magnetoelectronics; spintronics: devices
exploiting spin polarized transport or integrated magnetic fields}

\abstract{ We consider the magnetic response of a two-dimensional
electron gas (2DEG) with a spin-orbit interaction to a
long-wave-length electromagnetic excitation. We observe that the
transverse electric field creates spin polarization perpendicular
to the 2DEG plane.  The effect is more prominent in clean systems
with resolved spin-orbit-split subbands, and reaches maximum when
the frequency of the wave matches the subband splitting at the
Fermi momentum. The relation of this effect to the spin-Hall
effect is discussed.}

\begin{document}

\maketitle

Spintronics is a fast developing field of research. While the
spin-orbit effects in semiconductors have been discussed for a
long time~\cite{DyakonovPerel,aronov_lyanda,Edelstein}, recently
discovered spin-Hall effect~\cite{Sinova,Murakami} has generated a
lot of interest in applying spin-orbit related effects to
spintronics~\cite{murakami-2005-45}. The promise of the spin-Hall
effect is in the possibility of generation of non-equilibrium spin
polarization by means of a DC electric field.  However, after some
discussions it was concluded that in two-dimensional electron
gases with the Rashba and the Dresselhaus spin-orbit interactions,
the spin-Hall effect vanishes for constant and homogeneous
electric field~\cite{Inoue,Mishchenko,Dimitrova,chalaev-2005-71}.
At finite frequencies the spin-Hall effect is
non-zero~\cite{Mishchenko}.

While for the homogeneous spin-Hall effect (uniform applied electric field) the
out-of-plane spin polarization is expected to accumulate only at the edges of
the sample~\cite{Mishchenko}, there is an alternative possibility that we
propose to explore here: To create an out-of-plane inhomogeneous spin density
{\em in the bulk} in response to a spatially-modulated field.  Such bulk
accumulation is free from the uncertainties associated with the charge and spin
transport near the sample boundaries, and thus may provide an unambiguous
method to detect the spin-Hall effect.
\begin{figure}
\center{\includegraphics[width=0.9\columnwidth]{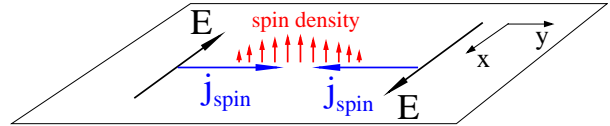}}
\caption[]{\label{fig:motiv1} The ``naive'' motivation: Spin
accumulation due to converging spin currents. }
\end{figure}

A  possible realization is shown in Fig.~\ref{fig:motiv1}, and
corresponds to a {\em transverse} electric field $E_x$ changing in
the $y$-direction.  In the presence of the spin-Hall effect, the
spin currents are expected to flow towards each other, with the
spin density accumulating in the middle. From the Maxwell
equation, ${\rm rot} \vecbf E = -(1/c)\dot{\vecbf B}$, a
transverse electric field can only exist together with the
time-dependent magnetic field. To create the situation shown in
Fig.~\ref{fig:motiv1} we therefore consider a linearly polarized
in-plane microwave field $\vecbf A = \vecbf e_x A_0 \exp(iq_y
y-i\Omega t)$; $\vecbf E = E_x \vecbf e_x  = \vecbf e_x (A_0/c) i
\Omega \exp(iq_y y-i\Omega t)$; $\vecbf B = B_z \vecbf e_z =
-\vecbf e_z A_0 i q_y \exp(iq_y y-i\Omega t)$; and $\Omega =
c|\vecbf q| = cq_y$.  There is naturally a spin response in the
$z$-direction due to the $B_z$ magnetic field (the Zeeman effect;
this is similar to the response to an in-plane field considered,
e.g., in Refs.~\cite{erlingsson-2005-71,shekhter-2005-71}).
However, in addition to this pure {\em spin} effect, we also find
an {\em orbital} contribution, $s_z^{\rm orbital} \sim {\rm rot}
\vecbf E$, which can be attributed to the spin-Hall effect. (Since
${\rm rot} \vecbf E \sim \Omega B_z$ this response can be
conveniently interpreted as an orbital contribution to the spin
susceptibility.) The response of the in-plane polarization to the
transverse electric field at arbitrary values of $\vecbf q$ was
considered for the clean case in Ref.~\cite{rashba-2005}.

We consider the Rashba model of a 2DEG with the Hamiltonian
\begin{equation}
H= \frac{\left(\vecbf p-\frac{e}{c}\,\vecbf A\right)^2}{2m^*} +
\alpha_{\rm R} \vecbf\eta \left(\vecbf p - \frac{e}{c}\,\vecbf
A\right) -\frac{1}{2}g\mu_{\rm B} \vecbf B \vecbf \sigma + V_{\rm
disorder}\ .
\end{equation}
where $\vecbf\eta \equiv \vecbf z \times \vecbf\sigma =
(-\sigma_y,\sigma_x,0)$ and $m^*$ is the effective electron band
mass. The model (with $\vecbf A=0$ and $V_{\rm disorder}=0$) has
two energy bands. The eigenfunctions of the two bands $\beta =
\pm$ are given by
\begin{eqnarray}
|u_{\pm,p}\rangle = \frac{1}{\sqrt{2}}\left(\ket{\uparrow} \mp i
e^{i\phi_p}\ket{\downarrow}\right)\ ,
\end{eqnarray}
where $\tan\phi_p\equiv p_y/p_x$ and the energies are given by
$\epsilon^{\pm}(\vecbf{p}) = \frac{\vecbf{p}^2}{2m^*}\pm
\Delta_p$, where $\Delta_p \equiv \alpha_{\rm R} |\vecbf{p}|$.

One should distinguish between the clean and dirty regimes. a) In
the clean regime (weak disorder or strong SOI) $\Delta_F \equiv
\alpha_{\rm R}\,p_F \gg \tau^{-1}$, where $\tau$ is the elastic
mean free time. The two spin-orbit split bands are, thus, well
defined. b) In the dirty regime (strong disorder or weak SOI)
$\Delta_F \ll \tau^{-1}$. The band splitting is not resolved. This
is the regime where the Dyakonov-Perel relaxation
applies~\cite{DyakonovPerel}. One can also define a ``super clean"
regime $1/\tau < m^* \alpha_{\rm R}^2$, which, however, appears to
be irrelevant experimentally at this time and will not be
considered here.

In this work we focus primarily on the clean regime, although our
diagrammatic derivation holds also in the dirty regime. Depending
on the frequency of the applied field one can then probe the {\em
resonant} ($\Omega\approx 2\Delta_F$) and {\em off-resonant}
($|\Omega-2\Delta_F|\gg 1/\tau$, e.g. $\Omega \ll 2\Delta_F$)
responses of the spin system.  Before rigorously deriving our
results diagrammatically, we provide qualitative derivation of
spin responses in these regimes.

In the low-frequency (off-resonant) limit, we consider slow (both
in time and space) components of the single-electron density
matrix $f_{\beta_1,\beta_2}(\vecbf{p}_1,\vecbf{p}_2)$. Here
$\beta_1$ and $\beta_2$ are the band indexes while $\vecbf{p}_1$
and $\vecbf{p}_2$ are the electron momenta. Introducing
$\vecbf{p}=(\vecbf{p}_1+\vecbf{p}_2)/2$ and $\vecbf{q}=\vecbf{p}_1
- \vecbf{p}_2$ we note that slowness in space means that $|\vecbf
q|$ is small. When the spin-orbit splitting, $\Delta_p$, is large,
only the band-diagonal elements
$f_\beta(\vecbf{p}_1,\vecbf{p}_2)\equiv
f_{\beta,\beta}(\vecbf{p}_1,\vecbf{p}_2)$ can be slow. Indeed the
band-off-diagonal elements either oscillate in time with frequency
$\sim \Delta_p$ or oscillate in space with $|\vecbf{q}|\sim
\Delta_p/v_F$. Consider the density matrix of band $\beta$ in the
coordinate representation:
\begin{eqnarray}
&&f_\beta (\vecbf{r},\vecbf{r}')=\nonumber\\
&&\sum_{\vecbf{p}_1,\vecbf{p}_2}\,f_\beta(\vecbf{p}_1,\vecbf{p}_2)\,
e^{(i\vecbf{p}_1 \vecbf{r} - i \vecbf{p}_2 \vecbf{r}')}\,
\ket{u_{\beta,\vecbf{p}_1}}\bra{u_{\beta,\vecbf{p}_2}} \ .
\end{eqnarray}
We calculate the $z$-component of the spin density (cf.
Ref.~\cite{Culcer}):
\begin{eqnarray}
s_z(\vecbf{r}) &=& \frac{1}{2}\sum_\beta{\rm Tr}\left[ \sigma_z
f_\beta(\vecbf{r},\vecbf{r})\right]\nonumber\\ &\approx&
\frac{1}{2}\sum_{\beta,\vecbf{p},\vecbf{q}}
\,f_{\vecbf{p},\vecbf{q},\beta}\, e^{i\vecbf{q}\vecbf{r}}
\bra{u_{\beta,\vecbf{p}}}\sigma_z\,\vecbf{q}\cdot
\nabla_{\vecbf{p}}
\,\ket{u_{\beta,\vecbf{p}}}\nonumber\\
&=& \frac{1}{2} \sum_{\beta,\vecbf{p},\vecbf{q}}
\,f_{\vecbf{p},\vecbf{q},\beta}\, e^{i\vecbf{q}\vecbf{r}}
\left(-i\vecbf{q}\cdot \nabla_{\vecbf p}\phi_p/2\right)\ .
\end{eqnarray}
Note that the contributions of both bands add up. For $\vecbf{q}=q_y
\vecbf{e_y}$ and using $\nabla_{p_y}\phi_p =\cos\phi_p/p$ we
obtain $s_z(\vecbf{q}) \propto i q_y \sum_{\beta,\vecbf{p}}
\,f_{\vecbf{p},\vecbf{q},\beta} \cos\phi_p /p$. From the usual
Boltzmann equation we know that under the influence of the
electric field $\vecbf{E}=E_x \vecbf{e_x}$ the correction to the
distribution function satisfies $\sum_{\vecbf{p}}
\,f_{\vecbf{p},\vecbf{q},\beta} \cos\phi_p \propto \nu e E_x v_F
\tau$, where $\nu$ is the density of states and $\tau$ is the
momentum relaxation time. Thus we finally obtain $s_z(\vecbf{q})
\propto i q_y e E_x \tau$ (we have used $p_F=m^* v_F$, $\nu \propto m^*$
and the fact that everything happens near the Fermi energy).

When the excitation frequency approximately matches the spin
subband splitting (the resonant case), the generation of the
transverse spin polarization can be understood based on the Bloch
equations, familiar from the theory of the nuclear and electron
spin resonance~\cite{Slichter}. The spin-orbit coupling can be
represented as a momentum dependent magnetic field $\vecbf b_0$
acting on the electrons with the Hamiltonian $H_{SO}=\vecbf b_0
\vecbf \sigma$, where $\vecbf b_0 = \Delta_p (p_y/p,-p_x/p)$. The
spin-dependent part of the interaction of the EM wave with
electrons can be recast in the following form, $H_{int} = {\vecbf
b_1 \vecbf \sigma}$ where $\vecbf b_1$ is a small oscillating
magnetic field directed along the $y$-axis, $\vecbf
b_1=\alpha_{\rm R} (e/c)A_0\vecbf e_y\,e^{iq_y y-i\Omega t}$. The
single-electron states $\ket{u_{-,\vecbf p}}$ and
$\ket{u_{+,\vecbf p+\vecbf q}}$ that are connected by the AC field
$\vecbf b_1$ {\em do not} have the same momentum, and thus their
energy difference is given by $\delta E_p = \epsilon^+({\bf p+q}) -
\epsilon^-({\bf p})\approx 2\Delta_p + p_y q_y/m^*$. For each such
pair of states, provided that, without the driving,
$\ket{u_{-,\vecbf p}}$ is occupied while $\ket{u_{+,\vecbf
p+\vecbf q}}$ is not, the usual Bloch dynamics of the electron
spin takes place. That is $\ket{u_{-,\vecbf p}}$ plays the role of
the state $\ket{\uparrow}$ along the direction of $\vecbf b_0$,
while $\ket{u_{+,\vecbf p+\vecbf q}}$ plays the role of
$\ket{\downarrow}$. Only the part of $\vecbf b_1$ perpendicular to
$\vecbf b_0$ is important, $|\vecbf b_{1,\perp}| = |\vecbf
b_1|\sin\phi_p$. In the linear response regime the driving
produces the out-of-plane polarization
\begin{equation}
\label{eq:single_spin_polarization}
\langle \sigma_{z,\vecbf p}\rangle
\sim {\rm Re}\left[ \frac{ie^{-i\Omega t} |\vecbf b_1|
\sin\phi_p}{[i/T_2 + (\Omega - \delta E_p)]}\right]\ ,
\end{equation}
where $T_2$ is the dephasing time. $T_2\sim \tau$ in the clean limit. The total
spin polarization is given by $s_z = L^{-2} \sum_{\bf p} \langle
\sigma_{z,\vecbf p}\rangle n_F(\epsilon^-_{{\bf p}})[1 - n_F(\epsilon^+_{\bf
p+q})]$. The summation is effectively restricted to the ring defined by $p_F +
\alpha_{\rm R} m^* >p > p_F - \alpha_{\rm R}m^*$.
(Note that in the limit $\vecbf q=0$ the
spin polarization vanishes, since $\langle \sigma_{z,\vecbf p}\rangle \propto
\sin\phi_p$, i.e. it has only a $p$-component. This actually means that that
there is a resonant spin current $j^z_y$ flowing in the
system~\cite{Mishchenko}.)   Restricting ourselves to the experimentally
relevant ``not-super-clean'' regime $1/\tau >\alpha_{\rm R}^2 m^*$, all states in
the ring $p_F + \alpha_{\rm R} m^* >p
> p_F - \alpha_{\rm R}m^*$ are ``in resonance'' simultaneously
(cf.~(\ref{eq:single_spin_polarization})) when $\Omega \sim
2\Delta_F$. Assuming also $v_F |\vecbf q| < 1/\tau$ we expand the
denominator of (\ref{eq:single_spin_polarization}) and after
summation obtain
\begin{equation}
\label{eq:qualitative_spin_polarization} s_z\sim {\rm Re}\left[
\frac{i e^{-i\Omega t}(e/c)A_0 q \Delta_F^2}{[i/T_2 + (\Omega -
2\Delta_F)]^2}\right]\ .
\end{equation}
Notice the strongly resonant (Fano shape) nature of this
contribution. A very similar result also obtains for the case of
the Dresselhaus spin-orbit interaction.  Interestingly, by noting
that $A_0/c = E_x/(i\Omega)$ and replacing $\Omega \rightarrow
i/\tau$ we can approximately recover the low-frequency spin
accumulation from the resonant result.  This indicates the close
connection between these two effects.

A resonant generation of out-of-plane spin polarization was also
recently proposed in Ref~\cite{duckheim-2006-2}. In that work, the
polarization appeared due to explicit symmetry breaking by a
strong constant magnetic field applied in the 2DEG plane, while in
our case the effect is due to the finite value of $\vecbf q$.

{\it Diagrammatic derivation.} We now derive these results in a
rigorous manner. We follow the route of
Refs.~\cite{Burkov,Mishchenko} and derive, first, the kinetic
(diffusion) equations governing the dynamics of the charge and
spin densities. We use the Keldysh technique. For introduction see
Ref.~\cite{RammerSmith}. Although such derivations are
standard~\cite{RammerSmith}, for completeness, we provide here the
main elements.

We consider only s-wave disorder scattering, that is $V_{\rm
disorder} = \sum_k u\delta(\vecbf r - \vecbf r_k)$ where $\vecbf
r_k$ are random locations with the average density $n_{\rm imp}$.
We employ the linear response: $H=H_0+H_1$ where
\begin{equation}
H_0= \frac{\vecbf p^2}{2m^*} + \alpha_{\rm R} \vecbf\eta \vecbf p +
V_{\rm disorder}\ ,
\end{equation}
and
\begin{eqnarray}
H_1 &=& -\frac{e}{2c}\,\left\{\vecbf v , \vecbf A\right\}_+ + e\phi
-\frac{1}{2}g\mu_{\rm B} \vecbf B \vecbf \sigma \nonumber\\&=&
-\frac{e}{2c}\,\left\{\left(\frac{\vecbf p}{m^*}+\alpha_{\rm
R}\vecbf\eta\right),\vecbf A\right\}_{+} + \hat\Phi\ .
\end{eqnarray}
To make the calculation gauge invariant we have added a scalar potential
perturbation, $\phi$. We neglect the $\vecbf A^2$ contribution and
introduce, for convenience, a 4-potential $\hat\Phi \equiv e\phi
\sigma_0 - (1/2)g\mu_{\rm B} \vecbf B \vecbf \sigma$.

\begin{figure}
\center{\includegraphics[width=0.85\columnwidth]{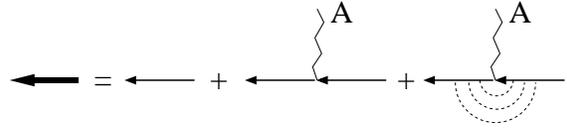}}
\caption[]{\label{fig:Diagramm}Diagrammatic representation of
Eq.~\ref{Eq:G_1}. The dashed lines represents the sum of impurity
``rainbow" diagrams (Diffuson) with minimum one impurity line.}
\end{figure}

In order to account properly for the vertex
corrections~\cite{Inoue,Dimitrova} we perform the linear response
expansion starting from the Dyson equation which reads $
(i\partial/\partial t - H)*G = \hat 1 + \Sigma*G \ , $ where *
stands for convolution. The self-energy is given by the
self-consistent Born approximation.
\begin{equation}
\label{Eq:SCBorn_Coordinates} \Sigma(\vecbf r_1,\vecbf
r_2,t_1,t_2) = n u^2 \delta(\vecbf r_1 - \vecbf r_2) G(\vecbf
r_1,\vecbf r_2,t_1,t_2)\ .
\end{equation}
Using $H=H_0+H_1$, $G=G_0+G_1$, and $\Sigma = \Sigma_0+\Sigma_1$
we obtain
\begin{equation}
\label{Eq:G_1} G_1 = G_0*(H_1+\Sigma_1)*G_0\ .
\end{equation}
The inclusion of $\Sigma_1$ is equivalent to accounting for the
vertex corrections. This can be seen from Fig.~\ref{fig:Diagramm}.

The zeroth order in $\vecbf A$ and $\hat\Phi$ Green's functions
and self energy reflect the standard disorder broadening. One
obtains $\Sigma_0^R = - i/(2\tau)$, where $\tau^{-1}=2\pi n_{\rm
imp} u^2 \nu$ and $\nu = m^*/(2\pi\hbar^2)$ is the density of states
(per spin) without the spin-orbit coupling. The zeroth order (in
$\vecbf A$ and $\hat\Phi$) disorder-broadened Green's functions
read
\begin{equation}
G_0^R = \left(\frac{1}{2}+\frac{1}{2}\frac{\vecbf\eta \vecbf
p}{p}\right)G_0^{R+} +
\left(\frac{1}{2}-\frac{1}{2}\frac{\vecbf\eta \vecbf
p}{p}\right)G_0^{R-}\ ,
\end{equation}
where $G_0^{R\pm}(\vecbf p,\omega)\equiv \left(\omega -
\epsilon^{\pm}(\vecbf p)+i/(2\tau)\right)^{-1}$. In equilibrium
$G^K_0 = h(\omega)\,(G^R_0-G^A_0)$ and $\Sigma^K_0 =
h(\omega)\,(\Sigma^R_0-\Sigma^A_0)$, where $h(\omega) \equiv
\tanh\frac{\omega-E_F}{2T}$.

Our goal is to find $G_1$ with the self-consistency condition
\begin{equation}
\label{Eq:SCBorn_2} \Sigma_1(\vecbf q,\omega,\Omega) =
\frac{1}{m^*\tau} \int \frac{d^2 p}{(2\pi)^2}\,G_1(\vecbf p,\vecbf
q,\omega,\Omega)\ .
\end{equation}
Having $G_1$ one can calculate any single-particle quantity, i.e.,
density or current. Thus this approach is equivalent to the Kubo
formula with the external vertex (the responding observable) being
chosen later.

As usual in the linear response theory the Keldysh component
$G^K_1$ splits into two parts, $G^K_1=G^{K,I}_1+G^{K,II}_1$.
Accordingly does the self-energy,
$\Sigma^K_1=\Sigma^{K,I}_1+\Sigma^{K,II}_1$. The first part,
$G^{K,I}_1$, corresponds to the R-A combinations in the Kubo
formula, while $G^{K,II}_1$ stands for the R-R and A-A
combinations~\cite{RammerSmith}. The spin-charge density matrix is
given by $\hat \rho = \frac{n}{2} + \vecbf s \vecbf \sigma= \int
\frac{d\omega}{2\pi} \int \frac{d^2 p}{(2\pi)^2} \,[-i\, G_1^<] =
-\frac{i}{2} \int \frac{d\omega}{2\pi} \int \frac{d^2
p}{(2\pi)^2}(G_1^K - G_1^R + G_1^A) = \hat\rho^{I}+\hat\rho^{II}$,
where $\hat\rho^{I}=-\frac{i}{2} \int \frac{d\omega}{2\pi} \int
\frac{d^2 p}{(2\pi)^2}\,G_1^{K,I}$ and $\hat\rho^{II}$ represents
the rest. Here $n(\vecbf q,\Omega)$ is the charge density while
$\vecbf s(\vecbf q,\Omega)$ is the spin density. We obtain
\begin{eqnarray}
\label{Eq:GK1I} &&G^{K,I}_1(\vecbf p,\vecbf
q,\omega,\Omega)=G^R_0(\vecbf p + \vecbf q/2,\omega +
\Omega/2)\cdot\nonumber\\
&&\cdot\,\left[\left\{h(\omega+\Omega/2)-h(\omega-\Omega/2)\right\}
H_1+\Sigma^{K,I}_1(\vecbf q,\omega,\Omega)
\right]\cdot\nonumber\\&&\cdot\,\,G^A_0(\vecbf p - \vecbf
q/2,\omega - \Omega/2)\ .
\end{eqnarray}
and $G^{K,II}_1 =  G^R_1 \, h(\omega-\Omega/2) -
h(\omega+\Omega/2)\, G^A_1$, where the correction to the retarded
Green's function is given by $G^R_1(\vecbf p,\vecbf
q,\omega,\Omega) = G^R_0(\vecbf p + \vecbf q/2,\omega +\Omega/2)
\cdot \left[H_1(\vecbf p,\vecbf q,\omega,\Omega)+\Sigma^R_1(\vecbf
p,\vecbf q,\omega,\Omega) \right] \cdot G^R_0(\vecbf p - \vecbf
q/2,\omega -\Omega/2)$. The dot product is used here to indicate
the matrix multiplication.

Our calculations below apply both in the clean (but not
``super-clean'') regime, $m^*\alpha_{\rm R}^2<\tau^{-1}<\Delta_F$,
and in the dirty regime, $\tau^{-1}>\Delta_F$. In the
``super-clean'' regime, on the other hand, the integration over
$\omega$ in the window of width $\Omega$ around the Fermi energy
(this window is cut by the function
$h(\omega+\Omega/2)-h(\omega-\Omega/2)$ in (\ref{Eq:GK1I})) is not
smooth and more care has to be taken.

Integrating Eq.~(\ref{Eq:GK1I}) over $\vecbf p$ and $\omega$ we
obtain
\begin{equation} \label{Eq:kin_rho}
\frac{(1-I)}{\tau}\hat\rho^{I} = i\Omega \nu\tilde
I\left[\frac{e\{\vecbf v \vecbf A\}_+}{2c}-\hat\Phi\right]\ ,
\end{equation}
The functional $\tilde I$ is defined as
\begin{eqnarray}
&&\tilde I[X(\vecbf p)]=\nonumber\\&&= \frac{1}{m^*\tau}\int
\frac{d^2p}{(2\pi)^2}\, G^R_0(\vecbf p + \vecbf q/2,E_F
+\Omega/2)\cdot  X(\vecbf p) \cdot\nonumber\\&&\cdot\,
G^A_0(\vecbf p - \vecbf q/2,E_F -\Omega/2)\ ,
\end{eqnarray}
while $I$ is a $4\times 4$ matrix defined by its action on the
$4$-vector $\hat\rho$ as $I \hat\rho = \tilde I[\hat\rho]$ (we
just use the fact that $\hat\rho$ is independent of $\vecbf p$ to
represent the functional $\tilde I[\hat\rho]$ as a product of a
$4\times 4$ matrix $I$ and a vector $\hat\rho$).

The second contribution to the density, $\hat\rho^{II}$ is given
by
\begin{equation}
\hat\rho^{II}=-\nu \hat\Phi\ .
\end{equation}
Thus the total density follows from
\begin{equation} \label{Eq:kin_rho_total}
\frac{(1-I)}{\tau}(\hat\rho+\nu\hat\Phi) = i\Omega \nu\tilde
I\left[\frac{e\{\vecbf v \vecbf A\}_+}{2c}-\hat\Phi\right]\ .
\end{equation}
We rewrite this equation to have all the terms associated with the
perturbing fields on the RHS:
\begin{equation} \label{Eq:kin_rho_total_RHS}
\frac{(1-I)}{\tau}\hat\rho = i\Omega \nu\tilde
I\left[\frac{e\{\vecbf v \vecbf A\}_+}{2c}\right] +
\frac{[(1-i\Omega\tau) I-1]}{\tau}\,\nu \hat\Phi \ .
\end{equation}

We allow for arbitrary external frequency $\Omega$, including
$\Omega> \tau^{-1}$.  However, we limit ourselves to the
experimentally relevant regime $v_F |\vecbf{q}| \ll \tau^{-1}$.
(Recently the spin and charge response functions to the {\em
longitudinal} fields were calculated for arbitrary values of
$\vecbf q$ in Ref.~\cite{Pletyukhov}.) Expanding the RHS of
Eq.~(\ref{Eq:kin_rho_total_RHS}) in powers of $\vecbf q$ up to the
terms linear in $\vecbf q$ we obtain
\begin{eqnarray} \label{Eq:kin_RHS_expanded}
&&\frac{(1-I)}{\tau}\hat\rho =-\nu (e\vecbf E\cdot \vecbf
\eta)\left[\frac{\alpha_{\rm
R}b^2}{2a(a^2+b^2)}\right]\nonumber\\&&\phantom{=}- (i\vecbf q
\times e\vecbf E)\cdot \vecbf \sigma \,
\left[\frac{b^2(a^2-b^2)}{8\pi a(a^2+b^2)^2}\right]
\nonumber\\
&&\phantom{=}+ \frac{b^2}{4\tau(a^2+b^2)}\,\nu\,g\,\mu_{\rm B}\,
\left(\vecbf B_x\sigma_x + \vecbf B_y\sigma_y + 2\vecbf
B_z\sigma_z  \right)\ ,
\nonumber\\
\end{eqnarray}
where $\vecbf E = i\Omega \vecbf A - i\vecbf q \phi$ (we also
assumed $div \vecbf E =0$). We have introduced $a\equiv
1-i\Omega\tau$ and $b\equiv 2\Delta_F\tau$. Note that the magnetic
terms ($\propto \vecbf B$) are already of first order in $\vecbf
q$. The term in the RHS of (\ref{Eq:kin_RHS_expanded})
proportional to $i\vecbf q\times \vecbf E = rot \vecbf E$ will lead to
our central result. The $\vecbf q$ expansion of the inverse "diffuson"
in the LHS of Eq.~(\ref{Eq:kin_rho_total_RHS}) reads
\begin{eqnarray}\label{Eq:kin_LHS_expanded}
&&\frac{1-I}{\tau}\,\hat\rho=\nonumber\\&&=\frac{1}{\tau}\,
\left[1-\frac{1}{a}\right]\,\frac{n}{2} +
\frac{1}{\tau}\,\left[1-\frac{2a^2+b^2}{2a(a^2+b^2)}\right]\,(s_x
\sigma_x + s_y \sigma_y) \nonumber\\&&+
\frac{1}{\tau}\,\left[1-\frac{a}{a^2+b^2}\right]\,s_z \sigma_z -
\sigma_\beta D_{nm}^{\beta\gamma} (i q_n)(i q_m){\rm Tr}
\left[\frac{\hat \rho\sigma_\gamma}{2}\right] \nonumber\\&&-
\frac{\alpha_{\rm R} b^2}{4 a^2 (a^2 +b^2)}\, \{\vecbf \eta,i
\vecbf q\,\hat\rho\}_{+}\, -\frac{iab v_F}{2(a^2+b^2)^2}\,[\vecbf
\eta, i\vecbf
q\,\hat\rho]_{-}\ ,\nonumber\\
\end{eqnarray}
where $D_{nm}^{\beta\gamma}$ is the diffusion tensor. In the dirty
limit, $b\ll 1$, we obtain $D_{nm}^{\beta\gamma} \approx
(D/a^3)\delta_{nm}\delta_{\beta\gamma}$, where $D=v_F^2 \tau/2$ is
the diffusion coefficient. In the clean limit, $b\gg 1$, only the
diffusion coefficient in the charge channel is important and we
obtain $D_{nm}^{00}\approx(D/a^3)\delta_{nm}$.
Eqs.~(\ref{Eq:kin_RHS_expanded}) and (\ref{Eq:kin_LHS_expanded})
generalize the results of Refs.~\cite{Burkov,Mishchenko} for
arbitrary frequency $\Omega$ and field polarization. Expanding
these equations in the dirty limit, $b\ll 1$, and for $\Omega\tau
\ll 1$ we obtain the equations of Ref.~\cite{Burkov}. The two last
terms of Eq.~(\ref{Eq:kin_LHS_expanded}) taken in the limit
$\Omega\tau \ll 1$ reproduce the spin-charge and spin-spin
couplings of Ref.~\cite{Mishchenko}. More importantly, we
introduce a new driving term in these equations, i.e., the second
term in the RHS of Eq.~(\ref{Eq:kin_RHS_expanded}) proportional to
$rot \vecbf E$. This term was not appreciated in
Refs.~\cite{Burkov,Mishchenko} where only longitudinal driving
fields were considered.

We will now use these equations to
obtain the spin density. We calculate contributions of zeroth and
of first orders in $\vecbf q$, $\vecbf s = \vecbf s_0 + \vecbf
s_1$. The first order contribution further splits into the orbital
and the Zeeman parts, $\vecbf s_1 = \vecbf s_1^{\rm orbital} +
\vecbf s_1^{\rm Zeeman}$.

{\it Zeroth order in $\vecbf q$.} The zeroth order in $\vecbf q$
corresponds to a homogeneous field. In this case there is no
difference between the transverse and the longitudinal cases. Thus
we should reproduce known results. For the spin polarization we
obtain (cf. Ref.~\cite{raimondi:035340})
\begin{equation}
\label{Eq:Edelstein} {\vecbf s}_0 = \alpha_{\rm
R}\,e\,E_x\,\tau\,\nu\, {\vecbf e_y}\,
\frac{b^2}{b^2(2a-1)+2a^2(a-1)} \ .
\end{equation}
For $\Omega\tau \ll 1$ in the case of strong spin-orbit ($b\gg 1$)
and for $\Omega \ll \Delta_F^2 \tau$ for the case of weak
spin-orbit ($b\ll 1$) this gives ${\vecbf s}_0(\Omega\rightarrow
0)\approx\alpha_{\rm R}\,e\,E_x\,\tau\,\nu\,{\vecbf e_y}$. In
other words this result is obtained when $\Omega \tau_s\ll 1$,
where $\tau_s \equiv \max[\tau,1/(\Delta_F^2\tau)]$ is the spin
relaxation time. This is the well known
result~\cite{aronov_lyanda,Edelstein} meaning that there is an
in-plane spin polarization perpendicular to the applied electric
field. In the clean limit, $\Delta_F\tau \gg 1$, at $\Omega
\approx 2\Delta_F$ the polarization (\ref{Eq:Edelstein}) shows
resonance discussed in Ref.~\cite{shekhter-2005-71}.

{\it First (linear) order in $\vecbf q$.}  After some algebra we
obtain
\begin{eqnarray}\label{Eq:rho_1}
&&\vecbf s_1^{\rm orbital} =\nonumber\\&& -{\vecbf e_z}\,\frac{i e
E_x\,\tau\,q_y}{4\pi}\,\frac{b}{(b^2+a^2-a)}\,
\left[\frac{b}{2a} - \Gamma_0\frac{ab}{a^2+b^2}\right]\nonumber\\
&&={\vecbf e_z}\,\frac{m_e}{m^*}\, \nu \mu_{\rm B} B_z
\,\frac{b(1-a)}{(b^2+a^2-a)}\,
\left[\frac{b}{2a} - \Gamma_0\frac{ab}{a^2+b^2}\right]\ ,\nonumber\\
\end{eqnarray}
where $ \Gamma_0=\frac{2(a-1)(b^2+a^2)}{b^2(2a-1)+2a^2(a-1)}$. We
have used $q_y E_x = - \Omega B_z/c$ and $\frac{e}{4\pi c} =
\left(\frac{m_e}{m^*}\right)\,\nu\mu_{\rm B}$, where $\mu_{\rm
B}\equiv e/(2m_e c)$ and $m_e$ is the bare electron mass. Thus we
obtain the out-of-plane spin polarization. This is the main result
of this work. It follows from the second term in the RHS of
Eq.~(\ref{Eq:kin_RHS_expanded}). For $\Omega\tau_s \ll 1$
\begin{equation}\label{Eq:rho_1_Omegatau0}
\vecbf s_1^{\rm orbital}(\Omega) \approx -{\vecbf
e_z}\,\frac{i\,e\,E_x\,\tau\,q_y}{8\pi}= {\vecbf
e_z}\,\frac{m_e}{m^*}\, \nu \mu_{\rm B} B_z
\,\frac{i\Omega\tau}{2}\ .
\end{equation}
The out-of-plane polarization is much smaller than the in-plane
one given by Eq.~(\ref{Eq:Edelstein}). Indeed we observe that for
$\Omega\tau_s \ll 1$ their ratio is given by $|\vecbf{s}^{\rm
orbital}_1|/|\vecbf{s}_0| \sim |\vecbf q| l_{SO}$, where $l_{SO} =
\hbar/(m^*\alpha_{\rm R})$ is the spin-orbit length. The fact that
the result is purely imaginary means that the spin wave is phase
shifted by $\pi/2$ relative to the applied microwave. This is
qualitatively consistent with the naive picture of the
$z$-polarized spin currents proportional to the electric field
(see Fig.~\ref{fig:motiv1}). However, for quantitative analysis
one would need to use the correct continuity equation $-i\Omega\,
s^z + i q_y\, j_y^z=t^z$, where $t^z$ is the torque
density~\cite{Burkov}.

Near the resonance, $\Omega \sim 2\Delta_F$, in the clean limit,
$\Delta_F\tau \gg 1$, we obtain
\begin{eqnarray}
&&\vecbf s_1^{\rm orbital}\approx -{\vecbf e_z}\,\frac{m_e}{m^*}\,
\nu \mu_{\rm B}
B_z\times\nonumber\\&&\times\frac{\Delta_F^2}
{(\Omega-2\Delta_F+\frac{i}{2\tau})
(\Omega-2\Delta_F+\frac{3i}{4\tau})}\ .
\end{eqnarray}
Thus we obtain a double pole like in
Eq.~(\ref{eq:qualitative_spin_polarization}).

{\it Response to the Zeeman term.} The Zeeman response follows
from the last term in Eq.~(\ref{Eq:kin_RHS_expanded}) and is given
by
\begin{equation}
\label{eq:s_z_Fermi} \vecbf s_1^{\rm Zeeman}  ={\vecbf e_z}\,\nu
\mu_{\rm B} B_z \, \frac{g}{2}\,\frac{b^2}{b^2-(1-a)a} \ .
\end{equation}
In the limit of zero spin-orbit coupling, i.e., $b=0$, but finite
$\Omega$ we observe that $\vecbf s_1^{\rm Zeeman}$ vanishes. The
physical meaning of this result is that without the spin-orbit
coupling no spin flips are possible and, thus, the spin density
cannot react to a dynamical perturbation. The other limiting
procedure (first $\Omega \rightarrow 0$, then $b \rightarrow 0$)
gives the thermodynamic Pauli susceptibility.

\begin{figure}
\center{\includegraphics[width=0.85\columnwidth]{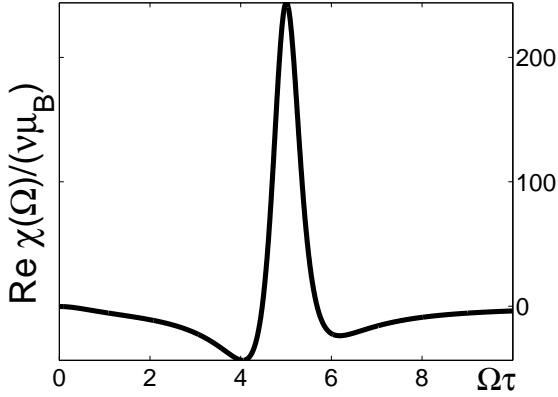}}
\caption[]{\label{fig:plot_real} Real part of the total spin
susceptibility $\chi\equiv s_z/B_z$ for $2\Delta_F\tau = 5$.
Parameters assumed as in GaAs: $m_e/m^* \approx 15$, $g\approx
-0.44$.}
\end{figure}
{\it Discussion.} In the clean limit $\Delta_F \tau \gg 1$ the
orbital out-of plane spin polarization (\ref{Eq:rho_1}) dominates
over $s_z^{\rm Zeeman}$ near resonance $\Omega \sim 2\Delta_F$
(although $s_z^{\rm Zeeman}$ is also resonant there). Also in the
clean limit the low frequency out-of-phase orbital response is
dominant, i.e., ${\rm Im} [s_{1z}^{\rm orbital}] \gg {\rm Im}
[s_{1z}^{\rm Zeeman}]$ for $\Omega\tau \ll 1$. In the opposite
regime, $\Delta_F \tau \ll 1$, the ratio of absolute values of the
orbital and the Zeeman contributions becomes independent of
$\Delta_F$,
\begin{equation}
\frac{|s_{1z}^{\rm orbital}|}{|s_{1z}^{\rm Zeeman}|}\approx
\frac{1}{g}\,\frac{m_e}{m^*}\,\frac{\Omega\tau}{\sqrt{1+\Omega^2\tau^2}}\ .
\end{equation}
If the prefactor $(1/g)(m_e/m^*)$ is large (as it happens, e.g., in GaAs)
the orbital contribution dominates at high enough frequencies.

\begin{figure}
\center{\includegraphics[width=0.85\columnwidth]{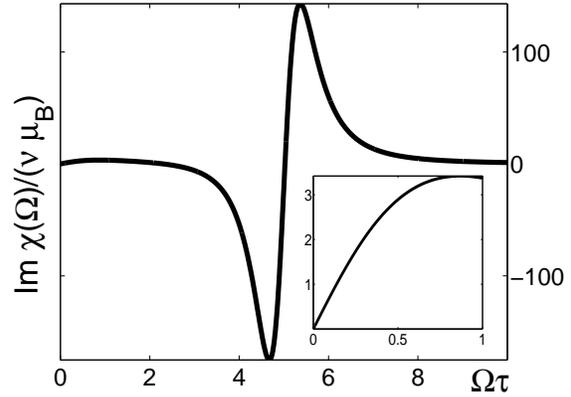}}
\caption[]{\label{fig:plot_imag} Imaginary part of the total spin
susceptibility $\chi\equiv s_z/B_z$ for $2\Delta_F\tau = 5$.
Inset: Blow-up of the imaginary part at $\Omega\tau < 1$.
Parameters assumed as in GaAs: $m_e/m^* \approx 15$, $g\approx
-0.44$.}
\end{figure}
It is convenient to measure the susceptibility in units of
$\nu\mu_{\rm B}$. In Figs.~\ref{fig:plot_real} and
\ref{fig:plot_imag} the resonant behavior at high frequency is
shown. For plotting we choose the clean regime, $2\Delta_F \tau =
5$. The orbital Fano-like contribution is clearly pronounced. The
linear in frequency dependence of the imaginary part at
$\Omega\tau < 1$ (shown in the inset in Fig.~\ref{fig:plot_imag})
corresponds to Eq.~(\ref{Eq:rho_1_Omegatau0}).

For numerical estimates we take the parameters of the 2DEG in
GaAs/AlGaAs heterostructures from Ref.~\cite{Zumbuehl_PRL02}. With
$m^* \sim 0.067 m_e \sim 0.7\cdot 10^{-31}$kg we obtain the
density of states $\nu = m^*/(2\pi\hbar^2) \sim 10^{36} $J$^{-1}$
m$^{-2}$. With $\mu_{\rm B}\sim 10^{-23}$ J/T we obtain
$\nu\mu_{\rm B} \sim 10^{13}$ T$^{-1}$m$^{-2} = 10^{5} $G$^{-1}
$cm$^{-2}$. With the mobility $\mu = e\tau/m^* \sim 25 $m$^2$/Vs
and the electron density $n_0 \sim 6\cdot 10^{15} $m$^{-2}$ we
obtain $\tau\sim 10^{-11}$s, $v_F \sim \sqrt{n_0/(m^*\nu)}\sim
3\cdot 10^5$m/s, and the mean free path $l =\tau v_F\sim 3 \mu $m.
The reported value for the spin-orbit length $l_{SO} \sim 4\mu m$.
Thus we are on the border between the Dyakonov-Perel (dirty)
regime and the strong spin-orbit regime so that $\Delta_F\tau =
l/l_{SO}\sim 1$.

The oscillating transverse electric field needed to generate the  bulk spin accumulation
described here can be created inside a microwave wave guide or a
strip-line resonator.  Assuming for the estimate a microwave field with $B_z\sim$ 1
Gauss, the spin polarization on resonance is about $2\cdot 10^7$ cm$^{-2}$.
Such polarization should be measurable with the magneto-optical spectroscopic
techniques~\cite{Crooker_PRB97,Kato_Science04}.  Alternatively, the transverse (circulating) electric
field could be generated with an AC magnet.

We acknowledge discussions with M. Freeman, A. Shekhter, M. Pletyukhov,  A. G.
Yashenkin, and S. Crooker. This work was supported by the DOE and by the CFN of
the DFG.

\bibliographystyle{mybst}
\bibliography{ref}

\end{document}